\documentclass[ floatfix,twocolumn]{revtex4}
\usepackage{amsmath}
\usepackage{color}
\usepackage{amssymb}
\usepackage{graphicx}
\usepackage[utf8]{inputenc}
\usepackage[english, russian]{babel}
\usepackage{textcomp}
\usepackage{color}
\usepackage{esint}

\begin{document}

\title{ 
Berry-Curvature Effect in Anomalous  Transport}

\author{V.P.Mineev}
\affiliation{Landau Institute for Theoretical Physics, 142432 Chernogolovka, Russia}

\begin{abstract}

A theory of the Berry-phase effect in anomalous transport in ferromagnets driven by statistical forces such as the gradient of temperature or chemical potential
has been developed by Di Xiao et al in Phys. Rev. Lett. {\bf 97}, 026603 (2006) based on a field-dependent density of states  generating  an additional contribution to the current density.
We present an alternative derivation of the Berry-phase effect in anomalous transport  in altermagnets and noncentrosymmertric metals based  on  quasiclassic 
matrix kinetic equation. Along with anomalous Hall effect an  anomalous heat flow and 
 anomalous thermalelectric currents are  derived. 
\end{abstract}
\date{\today}
\maketitle

\section{Introduction}

It is well known that the group velocity of a Bloch electron contains an anomalous term proportional to the mechanical force \cite{Niu1995,Xiao2010}, i.e.,
\begin{equation}
{v}_i^n=\frac{\partial\varepsilon^n_{\bf k}}{\partial k_i}+e {\Omega}^n_{ij}E_j,
\end{equation}
where ${ \Omega}^n_{ij}$ is the Berry curvature tensor of the $n$-th band with energy $\varepsilon^n_{\bf k}$. Corresponding Hall conductivity is
\begin{equation}
\sigma_{ij}=e^2\sum_n\int\frac{d^3{\bf k}}{(2\pi)^3} {\Omega}^n_{ij}n(\varepsilon_n).
\label{Hall}
\end{equation}
Here $n(\varepsilon_n)=\left \{\exp(\varepsilon_n-\mu)/T+1  \right \}^{-1}$ is the Fermi-Dirac distribution function. We put $\hbar=1$ here and everywhere below. 

The formula (\ref{Hall}) is derived based on the modification of the Bloch functions of electron   in presence of a mechanical force \cite{Niu1995,Xiao2010}.
The Berry phase also manifests in transport driven by a statistical forces induced by gradient of temperature or chemical potential.  
This problem has been solved in the paper \cite{Xiao2006} where were found the currents corresponding to  thermalelectric transport
and established the corresponding  Onsager relation between  transport coefficients connecting thermoelectric Hall currents and forces.  
This was made  by using the field-dependent density of states introduced in \cite{Xiao2005}  generating an additional contribution to the current density originating from
magnetic moment of an electron  wave packet \cite{Sundaram1999} , which has a finite spread in the phase space. 

The theory developed  in the papers  Ref. [1-5] operates with   spinless fermions. So, the  Berry curvature, corresponding magnetic moment of  wave packets and the magnetic field dependent density of states originate from the electron orbital   degrees of freedom. This allows to derive the anomalous transport properties in systems possessing this type of the Berry curvature.
Another situation is realised 
in altermagnetic  and also in  noncentrosymmetric metals where the electron band dispersion  possesses spin dependent band splitting characterised by the Berry curvature. In this case the theory of anomalous electric, thermal and thermalelectric transport can be developed based on another mechanism
making use  matrix quasiclassic kinetic equation.

The purpose of this article is to present the corresponding derivation.

 \section{Band splitting}

The Kramers degeneracy of electron states, known in metals, is lifted in ferromagnetic materials, as well as in noncentrosymmetric metals, that is, metals with a crystal structure without mirror symmetry. This leads to a splitting of the energy of the electron bands \cite{Roth1967,Mineev2005}.
 
 A new class of magnetic materials has been introduced recently in which sublattices with opposite spins can be coupled by proper or improper rotations, but cannot be coupled by translation or inversion. The electronic band structure in these materials consists of momentum-dependent spin-split bands, but, like antiferromagnets, lacks spontaneous bulk magnetization. The term ”altermagnets” has been proposed for materials of this type.
The approach allowing to determine the band structure has been implemented based on the so-called 
 exchange approximation including symmetry operations only in spin space which are compatible with crystal symmetry  \cite{Smejkal2022}.
Another approach to the altermagnets based on traditional symmetry classification of magnetic materials which includes both nonrelativistic and relativistic interactions has been developed in \cite{Mineev2025,MineevUFN}.

 Each electron energy  band in altermagnetic  or noncentrosymmetric metal  has the form \cite{Mineev2025,MineevUFN}
\begin{equation}
\label{H_0}
 \hat \varepsilon
 = \varepsilon\sigma_0+\mbox{\boldmath$\gamma$} 
   \cdot \mbox{\boldmath$\sigma$},
\end{equation}
where 
$\varepsilon=\varepsilon({\bf k})$ denotes the spin-independent part of the spectrum,
$\sigma_0$ is the unit $2\times 2$ matrix in the spin space, $\mbox{\boldmath$\sigma$}=(\sigma_x,\sigma_y,\sigma_z)$ are the Pauli matrices. 
The pseudovector 
\begin{equation}
\mbox{\boldmath$\gamma$}=\mbox{\boldmath$\gamma$}({\bf k})=\gamma_x({\bf k})\hat x+\gamma_y({\bf k})\hat y+\gamma_z({\bf k})\hat z
\end{equation}  
is periodic 
 in the reciprocal space, even in altermagnets  $\mbox{\boldmath$\gamma$}(-{\bf k})=\mbox{\boldmath$\gamma$}({\bf k})$ and      odd in noncentrosymmetric metals  
$\mbox{\boldmath$\gamma$}(-{\bf k})=-\mbox{\boldmath$\gamma$}({\bf k})$ function. It satisfies
$g\mbox{\boldmath$\gamma$}(g^{-1} {\bf k})=\mbox{\boldmath$\gamma$}({\bf k})$,
where $g$ is any symmetry operation in the  point group ${\cal G}$ of
the crystal.

The eigenvalues of the matrix (\ref{H_0}) are
\begin{equation}
    \varepsilon_{+}({\bf k})=\varepsilon+\gamma,~~~~~~~~ \varepsilon_{-}({\bf k})=\varepsilon-\gamma,
\label{e3}
\end{equation}
where $\gamma=|\mbox{\boldmath$\gamma$}({\bf k})|$.
The corresponding eigenfunctions are given by
\begin{eqnarray}
\Psi^+_\sigma({\bf k})=\frac{1}{\sqrt{2\gamma(\gamma+\gamma_z)}}\left (\begin{array} {c}
\gamma+\gamma_z\\
\gamma_+
\end{array}\right),\nonumber\\
~~~~~~~~~~~~\Psi^-_\sigma({\bf k})=\frac{1}{\sqrt{2\gamma(\gamma+\gamma_z)}}
\left(\begin{array} {c}
-\gamma_-\\
\gamma+\gamma_z
\end{array}\right),
\end{eqnarray}
where $\gamma_\pm=\gamma_x\pm i\gamma_y$.
The eigenfunctions obey the orthogonality conditions
\begin{equation}
\Psi^{\alpha\star}_\sigma({\bf k})\Psi^\beta_\sigma({\bf k})=\delta_{\alpha\beta},~~~~~~~
\Psi^\alpha_{\sigma_1}({\bf k})\Psi^{\alpha\star}_{\sigma_2}({\bf k})=\delta_{\sigma_1\sigma_2}.
\label{ort}
\end{equation}
Here, a summation over the repeating  spin $\sigma=\uparrow,\downarrow$
or band $\alpha=+,-$ indices is implied. 

There are two Fermi surfaces determined by the equations
\begin{equation}
\label{e4}
    \varepsilon_{+}({\bf k})=\mu,~~~~~~~~~~~~\varepsilon_{-}({\bf k})=\mu
\end{equation}
and the Fermi velocities are given by the derivatives
\begin{equation}
{\bf v}_{+}=\frac{\partial\varepsilon_{+}({\bf k})}{\partial {\bf k}},~~~~~~~~~~~~~~{\bf v}_{-}=\frac{\partial\varepsilon_{-}({\bf k})}{\partial {\bf k}}~
\end{equation}
taken at points ${\bf k}$ lying on the Fermi surfaces.

Each band acquires  momentum dependent spin splitting. The Berry curvatures of splitted subbands are
\begin{equation}
\Omega_{ij}^{\pm}({\bf k})=-2Im\frac{\partial\Psi_\sigma^{\star\pm}}{\partial k_i}\frac{\partial\Psi_\sigma^{\pm}}{\partial k_i}=\mp e_{\alpha\beta\gamma}\frac{\gamma_\alpha}{2\gamma^3}\left (\frac{\partial\gamma_\beta}{\partial k_i}\frac{\partial\gamma_\gamma}{\partial k_j}  \right )
\label{Omega}
\end{equation}
Thus, the Hall conductivity is
\begin{eqnarray}
\sigma_{ij}=e^2\int\frac{d^3{\bf k}}{(2\pi)^3} \left [{\Omega}^+_{ij}n(\varepsilon_+) +{\Omega}^-_{ij}n(\varepsilon_-)\right]\nonumber\\
=e^2\int\frac{d^3{\bf k}}{(2\pi)^3} {\Omega}^+_{ij}\left [n(\varepsilon_+) -n(\varepsilon_-\right].
\label{Hall1}
\end{eqnarray}

Let us now calculate all currents using the Boltzmann equation.

\section{Kinetic equation approach}

The quasiclassic Boltzmann  equation for  a matrix distribution function  $\hat n({\bf k},{\bf r},t)$ of quasiparticles of a Fermi-liquid with spin $1/2$  in magnetic field has been derived by V.P.Silin
\cite{Silin1957}.  It is also applicable to the description of gas of electrons  filling the states in the spin split conduction band in altermagnets or in noncentrosymmetric metals.
It has the following form
\begin{eqnarray}
\frac{\partial \hat n}{\partial t}+\frac{1}{2}\left ( \frac{\partial\hat \varepsilon}{\partial {\bf k}}
\frac{\partial\hat n}{\partial {\bf r}} +\frac{\partial\hat n}{\partial {\bf r}} \frac{\partial\hat \varepsilon}{\partial {\bf k}}  \right )~~~~~~~~\nonumber\\-
\frac{1}{2}\left ( \frac{\partial\hat \varepsilon}{\partial {\bf r}}
\frac{\partial\hat n}{\partial {\bf k}} +\frac{\partial\hat n}{\partial {\bf k}} \frac{\partial\hat \varepsilon}{\partial {\bf r}}  \right )-i[\hat\varepsilon,\hat n]=
\hat I,
\label{Silin}
\end{eqnarray}
where $[\hat\varepsilon,\hat n]$ is the commutator.  The  collision integral in the rhs determines the relaxation processes.

The matrix of the equilibrium electron distribution function is
\begin{equation}
\hat n^{(0)}
=\frac{n({\varepsilon}_+)+n({\varepsilon}_-)}{2}\sigma_0+\frac{n({\varepsilon}_+)-n({\varepsilon}_-)}{2\gamma} 
\mbox{\boldmath$\gamma$}\cdot \mbox{\boldmath$\sigma$},
\label{eqv}
\end{equation}
where 
\begin{equation}
n({\varepsilon})=\frac{1}{\exp\left(\frac{\varepsilon-\mu}{T}\right)+1}
\end{equation}
is the Fermi function.

The Hermitian matrices of the non-equilibrium distribution functions in the band and spin representations  related by
\begin{equation}
f_{\alpha\beta}({\bf k})=\Psi^{\alpha\star}_{\sigma_1}({\bf k})n_{\sigma_1\sigma_2}\Psi^{\beta}_{\sigma_2}({\bf k}).
\label{f}
\end{equation}
In the band representation the equilibrium distribution function  is given by the diagonal matrix
\begin{equation}
n^{(0)}_{\alpha\beta}=\Psi^{\alpha\star}_{\sigma_1}({\bf k})n^{(0)}_{\sigma_1\sigma_2}\Psi^{\beta}_{\sigma_2}({\bf k})=\left (\begin{array} {cc} n({\varepsilon}_+)&0\\0&n({\varepsilon}_-)  \end{array}\right).
\end{equation}
The derivative of the energy matrix (\ref{H_0}) in  band representation is
\begin{eqnarray}
\Psi_{\sigma_1}^{\star\alpha}\frac{\partial\varepsilon_{\sigma_1\sigma_2}}{\partial{\bf k}}\Psi_{\sigma_2}^{\beta}= \left(\begin{array} {cc} {\bf v}_+
&{\bf w}_\pm(\varepsilon_--\varepsilon_+) \\
{\bf w}_\mp(\varepsilon_+-\varepsilon_-)&{\bf v}_- \end{array}\right).
\end{eqnarray}
Here, 
\begin{equation}
{\bf w}_{\pm}({\bf k})=
\Psi^{+\star}_{\sigma}({\bf k})\frac{\partial \Psi^{-}_{\sigma}({\bf k})}{\partial{\bf k}},
\end{equation}
\begin{equation}
{\bf w}_{\mp}=-{\bf w}_{\pm}^\star.~~~~~~~~~~~~~~~~~
\label{vel}
\end{equation}

In  presence of an electric field and gradients of temperature and chemical potential  the matrix  kinetic equation in the band representation 
 for the
stationary deviation of the distribution function from equilibrium distribution
\begin{equation}
g_{\alpha\beta}({\bf k})=f_{\alpha\beta}({\bf k})-n^{(0)}_{\alpha\beta}
\end{equation}
 is
\begin{eqnarray}
 \left(
\begin{array} {cc}{\bf v}_{+}\left (\frac{\partial n_+}{\partial\varepsilon_+}e{\bf E}+\nabla n_+\right)&-{\bf w}_{\pm}{\bf N}\\
{\bf w}_{\mp}{\bf N} &  {\bf v}_-\left (\frac{ \partial n_-}{\partial\varepsilon_-}e{\bf E}+\nabla n_-\right)
      \end{array}\right)\nonumber\\
+               
 \left(
\begin{array} {cc}0&ig_{\pm}(\varepsilon_--\varepsilon_+)\\
ig_{\mp}(\varepsilon_+-\varepsilon_-)&0
 \end{array}\right)
 = \left(
\begin{array}{cc}I_+&I_{\pm}\\I_\mp&I_- \end{array}\right).
 \label{eqv2}
\end{eqnarray}
Here,
\begin{equation}
{\bf N}=(n_+-n_-)e{\bf E}+(\varepsilon_+-\varepsilon_-)\frac{1}{2}\nabla(n_++n_-) ,
\label{N}
\end{equation}
\begin{equation}
\nabla n_\lambda=-\frac{\partial n_\lambda}{\partial\varepsilon_\lambda}\left (\nabla\mu+\frac{\xi_\lambda}{T}\nabla T\right ).
\end{equation}
We have used notations  $$n_\lambda=n(\varepsilon_\lambda), ~~\xi_\lambda=\varepsilon_\lambda-\mu, ~~\lambda=\pm.$$

 The electric  current density is determined by the following expression
 \begin{equation}
 {\bf j}=e\int \frac{d^3k}{(2\pi)^3}\frac{\partial \varepsilon_{\sigma\sigma_1}({\bf k})}{\partial {\bf k}}g_{\sigma_1\sigma}({\bf k},\omega).
 \end{equation}
Transforming it to the band representation we get
\begin{equation}
{\bf j}=e\int \frac{d^3k}{(2\pi)^3} \left \{ {\bf v}_+g_++{\bf v}_-g_-+({\bf w}_{\pm}g_{\mp}- {\bf w}_{\mp}g_{\pm})(\varepsilon_--\varepsilon_+)  \right \}.
\label{cur}
\end{equation}

The  density of heat current  is
 \begin{equation}
 {\bf q}=\frac{1}{2}\int \frac{d^3k}{(2\pi)^3}
 \left (\xi_{\sigma\sigma_1}
\frac{\partial \varepsilon_{\sigma_1\sigma_2}}{\partial {\bf  k}}+
 \frac{\partial \varepsilon_{\sigma\sigma_1}}{\partial {\bf  k}}\xi_{\sigma_1\sigma_2}\right )g_{\sigma_2\sigma}({\bf k}).
 \end{equation}

Transforming it to the band representation we obtain
\begin{eqnarray}
 {\bf q}=\int \frac{d^3k}{(2\pi)^3}
 \left \{\xi_+ {\bf v}_+g_+
 +\xi_-{\bf v}_-g_-\right.\nonumber\\
 \left.+\frac{1}{2}
 (\xi_++\xi_-)\left ({\bf w}_\pm g_\mp-{\bf w}_\mp g_\pm\right )(\varepsilon_--\varepsilon_+)  \right \}.
 \label{hcur}
 \end{eqnarray}
 
 When the band splitting strongly exceeds the typical scattering rate of quasiparticles
 \begin{equation}
 |\varepsilon_+-\varepsilon_-|\gg 1/\tau,
 \end{equation}
one can neglect by the off-diagonal terms in collision integral and put $I_\pm=I_\mp=0$ \cite{MineevUFN}.
Thus, the off-diagonal  matrix elements of distribution function are
\begin{eqnarray} 
g_\pm=i\frac{{\bf w}_{\pm}{\ \bf N}}{~\varepsilon_+-\varepsilon_-},\\
g_\mp=i \frac{{\bf w}_\mp{\bf N}}
{~\varepsilon_+-\varepsilon_-}.
\end{eqnarray}
Substituting these expression in Eqs.(\ref{cur}) and (\ref{hcur}) we obtain anomalous parts of current and heat current densities
\begin{equation}
j_i=e\int \frac{d^3k}{(2\pi)^3}( -2Im~w^{\pm}_iw^{\star\pm}_j)N_j,
\label{j}
\end{equation}
\begin{equation}
q_i= \frac{1}{2}\int \frac{d^3k}{(2\pi)^3}
 (\xi_++\xi_-)( -2Im~w^{\pm}_iw^{\star\pm}_j)N_j.
 \label{q}
\end{equation}
Here, we have used the relation given by Eq.(\ref{vel}).
One can check by direct calculation that
\begin{equation}
-2Im~w^{\pm}_iw^{\star\pm}_j=\Omega_{ij}^+({\bf k}),
\label{Omega_1}
\end{equation}
where $\Omega_{ij}^+({\bf k})$ is the tensor of the Berry curvature given by Eq.(\ref{Omega}).

After substitution in Eqs.(\ref{j}), (\ref{q}) expression(\ref{N}) we find that the electric current and the heat current have the following structure
\begin{eqnarray}
{\bf j}={\bf j}^E+{\bf j}^T,\\
{\bf q}={\bf q}^E +{\bf q}^T.
\end{eqnarray}
Here,
\begin{eqnarray}
j_i^E=e\int \frac{d^3k}{(2\pi)^3}\Omega_{ij}^+({\bf k})\left\{ (n_+-n_-)eE_j\right.\nonumber\\
\left.-\frac{1}{2}(\varepsilon_+-\varepsilon_-)\left(\frac{\partial n_+}{\partial\xi_+}+\frac{\partial n_-}{\partial\xi_-}\right)\nabla_j\mu\right\},
\label{jE}
\end{eqnarray}
\begin{eqnarray}
q_i^T=-\frac{1}{4T}\int \frac{d^3k}{(2\pi)^3}\Omega_{ij}^+({\bf k})(\xi_++\xi_-)(\varepsilon_+-\varepsilon_-)\nonumber\\
\times\left(\xi+\frac{\partial n_+}{\partial\xi_+}+\xi_-\frac{\partial n_-}{\partial\xi_-}\right)\nabla_jT,
\label{qT}
\end{eqnarray}
the thermal electric current is
\begin{eqnarray}
j_i^T=-\frac{e}{2T}\int \frac{d^3k}{(2\pi)^3}\left\{\Omega_{ij}^+(\varepsilon_+-\varepsilon_-)\right.\nonumber\\
\left.\times\left ( \xi_+ \frac{\partial n_+}{\partial \xi_+}+ \xi_- \frac{\partial n_-}{\partial \xi_-}\right )\right\}
\nabla_jT,
\label{jT}
\end{eqnarray}
the thermal electric heat flow is
\begin{eqnarray}
q_i^E=\frac{1}{2}\int \frac{d^3k}{(2\pi)^3}\left\{\Omega_{ij}^+(\xi_++\xi_-)~~~~~~~~~~~~~~~\right.\nonumber\\
\left.\times\left [(n_+-n_-)eE_j-\frac{1}{2}(\varepsilon_+-\varepsilon_-)\left  (\frac{\partial n_+}{\partial\varepsilon_+}+\frac{\partial n_-}{\partial\varepsilon_-}\right )\nabla_j\mu
\right ]\right\}.
\label{qE}
\end{eqnarray}
Under fulfilment the condition
\begin{equation}
\varepsilon_+-\varepsilon_-=2\gamma\ll\mu
\label{cond}
\end{equation}
the Eqs. (\ref{jE}) and (\ref{qT}) acquire more simple form
\begin{eqnarray}
j_i^E=\sigma_{ij}(E_j-\nabla_j\mu/e)~~~~~~~~~~~~~~~~~~~\nonumber\\=2e^2\int \frac{d^3k}{(2\pi)^3}\Omega_{ij}^+({\bf k})\gamma\frac{\partial n}{\partial\xi}(E_j-\nabla_j\mu/e),
\end{eqnarray}
\begin{equation}
q_i^T=-\kappa_{ij}\nabla T=-\frac{2}{T}\int \frac{d^3k}{(2\pi)^3}\Omega_{ij}^+({\bf k})\gamma\xi^2\frac{\partial n}{\partial\xi}\nabla_jT.
\end{equation}
After integration over $\xi$ expressions for conductivity and  thermal-conductivty are
\begin{equation}
\sigma_{ij}=-2e^2N_0\int \frac{d\Omega}{4\pi}\Omega_{ij}^+\gamma,
\end{equation}
\begin{equation}
\kappa_{ij}=-\frac{2\pi^2N_0T}{3}\int \frac{d\Omega}{4\pi}\Omega_{ij}^+\gamma.
\end{equation}
Thus, we come to equality
\begin{equation}
\kappa_{ij}=\frac{\pi^2T}{3e^2}\sigma_{ij}
\end{equation}
sometimes called as the Mott relation.

Under fulfilment the condition (\ref{cond})
the thermal electric current (\ref{jT})
\begin{equation}
j_i^T=\alpha^T_{ij}\nabla_jT
\end{equation}
and the heat flow (\ref{qE}) 
\begin{equation}
q_i^E=\alpha^E_{ij}(E_j-\nabla_j\mu/e)
\end{equation} 
are proportional to corresponding  thermodynamic forces with coefficients proportionality 
\begin{equation}
-T\alpha^T_{ij}=\alpha^E_{ij}=2e\int \frac{d^3k}{(2\pi)^3}\gamma\xi\frac{\partial n}{\partial \xi}\Omega_{ij}^+.
\end{equation}
related to each other according to the Onsager principle. Performing integration over $\xi$ we obtain
\begin{equation}
-T\alpha^T_{ij}=\alpha^E_{ij}=-\frac  {2\pi^2eN_0^\prime T^2}{3}\int \frac{d\Omega}{4\pi}\Omega_{ij}^+\gamma.
\end{equation}
Here, $N_0^\prime$ is the derivative of the density of states at the Fermi level.

\section{Berry curvature}

The band spectrum in noncentrosymmetric and altermagnet metals possesses momentum dependent band splitting described by the vector function  $\mbox{\boldmath$\gamma$}({\bf k})$. 
For instance,  in noncentrosymmetric metals CeRhSi$_3$ and LaRhSi$_3$ with tetragonal structure this function is \cite{Mineev2005}
\begin{equation}
\mbox{\boldmath$\gamma$}_{tetr}({\bf k})=c_1(k_x\hat y-k_y\hat x)+c_2k_xk_yk_z(k_x^2-k_y^2).
\label{NCS}
\end{equation}
The corresponding function in altermagnet CrSb with hexagonal structure is \cite{Mineev2026}
 \begin{equation}
\mbox{\boldmath$\gamma$}_{hex}({\bf k})=c_1\left[(k_x^2-k_y^2)\hat x-2k_xk_y\hat y\right ]+c_2k_zk_x(k_x^2-3k_y^2)\hat z.
\label{Aso}
\end{equation}
 Substituting  the Berry curvatures calculated  according to Eq.(\ref{Omega}) for 
$\mbox{\boldmath$\gamma$}_{tetr}({\bf k})$ and  $\mbox{\boldmath$\gamma$}_{hex}({\bf k})$ functions  
to any expression for anomalous currents is easy to check that all the corresponding  integrals are equal to zero due to symmetry properties of   $ \Omega_{ij}^+$({\bf k}).
However, in presence of magnetic field  along the symmetry axis of uniaxial crystals both   functions (\ref{NCS}) and (\ref{Aso})
include the Zeemann term 
\begin{equation} 
\mbox{\boldmath$\gamma$}({\bf k})\to\mbox{\boldmath$\gamma$}({\bf k})-\mu_BH\hat z
\end{equation}
and the integrals for  anomalous currents  depending on $ \Omega_{xy}^+$ acquire  finite values.

\smallskip

\section{Conclusion}

The derivation of  anomalous currents originating from the Berry curvature is presented based on  quasiclassic matrix kinetic equation applied to  noncentrosymmetric and altermagnet metals. 
Expressions for  anomalous heat flow and anomalous thermalelectric currents are established.  As it is in the case of the ordinary Hall effect the off-diagonal conductivity and thermal-conductivity expressed through the Berry curvature are related by the Mott relation. The cross coeffiicients for the thermal electric current and the thermal electric heat flow are expressed trough each other according to the Onsager symmetry principle.


\begin{thebibliography}{220}

\bibitem{Niu1995} Ming-Che Chang and Qian Niu, {\it Berry Phase, Hyperorbits, and the Hofstadter Spectrum}, Phys. Rev. Lett. {\bf 75}, 1348 {1995}.

\bibitem{Xiao2010}Di Xiao, Ming-Che Chang, Qian Niu, {\it Berry phase effects on electronic properties}, Rev. Mod.Phys. {\bf 82}, 1959 {2010}.

\bibitem{Xiao2006} Di Xiao, Yugui Yao, Zhong Fang, and Qian Niu, {\it Berry-Phase Effect in Anomalous Thermoelectric Transport}, Phys. Rev. Lett. {\bf 97}, 026603 {2006}.

\bibitem{Xiao2005}Di Xiao, Junren Shi, and Qian Niu, {\it  Berry Phase Correction to Electron Density of States in Solids}, Phys. Rev. Lett. {\bf 95,} 137204 (2005).

\bibitem{Sundaram1999} G. Sundaram and Qian Niu, {\it 
Wave-packet dynamics in slowly perturbed crystals: Gradient corrections and Berry-phase effects}, Phys.Rev.B {\bf 59}, 14915 (1999). 

\bibitem{Roth1967}L. M. Roth, S. H. Groves, and P. W. Wyatt, {\it Inversion asymmetry effects on oscillatory magnetoresistance in HgSe}, Phys. Rev. Lett. 
{\bf 19}, 576 (1967).

\bibitem{Mineev2005}V. P. Mineev and K. V. Samokhin, {\it De Haas-van Alphen effect in metals without inversion center }, Phys. Rev. B {\bf 72}, 212504 (2005).

\bibitem{Smejkal2022} L.\v{S}mejkal,  J. Sinova, T. Jungwirth, {\it Beyond Conventional Ferromagnetism and Antiferromagnetism: A Phase with Nonrelativistic Spin and Crystal Rotation Symmetry},  Phys. Rev.X {\bf 12}, 031042 (2022).

\bibitem{Mineev2025}V.P.Mineev,  {\it Altermagnetic and Noncentrosymmetric Metals}, JETP Letters {\bf 121}, 421 (2025).

\bibitem{MineevUFN}V.P.Mineev, {\it Toroid, altermagnetic, and noncentrosymmetric ordering in metals},
Physics - Uspekhi {\bf 68}, 1151 (2025).

\bibitem{Silin1957} V.P.Silin, {\it Oscillations of a Fermi-liquid in a magnetic field}, Zh.  Eksp.Teor.Fiz.  {\bf 33}, 1227 (1957) [Sov.  Phys. JETP {\bf 6}, 945 (1958)]. 


\bibitem{Mineev2026}V.P.Mineev, {\it Band splitting  in the altermagnet CrSb}, Phys. Rev. B  {\bf 114}, 065130 (2026).


\end{thebibliography}
\end{document}